\documentclass[12pt]{article}

\usepackage[totalheight = 23cm, totalwidth = 17cm]{geometry}
\usepackage{amssymb,amsmath,amsfonts,amsbsy,graphicx}

\def\mathbi#1{\textbf{\em #1}}

\newcommand{\fnl}{f_{\rm NL}}
\newcommand{\calO}{{\cal O}}
\newcommand{\calR}{{\cal R}}

\begin{document}

\begin{titlepage}

\begin{center}

\vspace*{-10ex}
\hspace*{\fill}
CERN-PH-TH/2011-199

\vskip 1.5cm

\Huge{Classical non-Gaussianity \\ from non-linear evolution \\ of curvature perturbations}

\vskip 1cm

\large{
Jinn-Ouk Gong$^{*}$\footnote{jinn-ouk.gong@cern.ch},
\hspace{0.2cm}
Jai-chan Hwang$^{\dag,\ddag}$\footnote{jchan@knu.ac.kr}
\hspace{0.2cm}\mbox{and}\hspace{0.2cm}
Hyerim Noh$^{\S}$\footnote{hr@kasi.re.kr}
\\
\vspace{0.5cm}
{\em
${}^*$ Theory Division, CERN
\\
CH-1211 Gen\`eve 23, Switzerland
\\
\vspace{0.2cm}
${}^\dag$ Department of Astronomy and Atmospheric Sciences
\\
Kyungpook National University
\\
Daegu 702-701, Republic of Korea
\\
\vspace{0.2cm}
${}^\ddag$ Korea Institute for Advanced Study
\\
Seoul 130-722, Republic of Korea
\\
\vspace{0.2cm}
${}^\S$ Korea Astronomy and Space Science Institute
\\
Daejeon 305-348, Republic of Korea}
}

\vskip 0.5cm

\today

\vskip 1.2cm

\end{center}

\begin{abstract}

We study the non-linear evolution of the curvature perturbations during matter dominated era. We show that regardless of the origin of the primordial perturbation, the Bardeen potential and curvature receive sizable contributions from the classical non-linear evolution effects, and quantify them exactly. On the super-horizon scales we have squeezed peak of the bispectrum with magnitude, in terms of the local non-linear parameters of Bardeen curvature, $1/6\leq\fnl\lesssim19/15$, and of Bardeen potential, $-1/4\leq\fnl\lesssim7/5$, depending on the configuration of momenta. On the sub-horizon scales the bispectrum show equilateral shape, and can serve as a potential probe of general relativity.

\end{abstract}

\end{titlepage}

\setcounter{page}{0}
\newpage
\setcounter{page}{1}

The ongoing observation programs have been bringing the era of precision cosmology. The combination of the Wilkinson Microwave Anisotropy Probe on the cosmic microwave background (CMB) and the Sloan Digital Sky Survey on large scale structure (LSS) results show that the primordial perturbation is characterized by a nearly flat power spectrum with almost perfect Gaussianity~\cite{Komatsu:2010fb}. This is consistent with the predictions of the paradigm of inflation~\cite{Guth:1980zm}, which provide the most successful mechanism for the causal generation of the primordial perturbation on cosmologically relevant scales. The upcoming surveys on the CMB and LSS, such as Planck, BigBOSS, Large Synoptic Survey Telescope and Euclid, will provide data even more precise enough to detect small deviations, if there are any, from what the simplest single field inflation model predicts. They are especially important as we can gain more information crucial for the primordial perturbation and the generation mechanism, thereby constraining the early universe and high energy physics~\cite{Lyth:2009zz}.

Among the possible deviations, non-Gaussianity has been receiving great interest for decades~\cite{Bartolo:2004if}. Usually, the simplest non-Gaussianity is parametrized by a dimensionless non-linear parameter $\fnl$ of the Bardeen curvature $\Phi$, the curvature perturbation in the zero-shear gauge (often termed the Newtonian gauge), as~\cite{Komatsu:2010fb,Komatsu:2001rj}
\begin{equation}\label{PhifNL}
\Phi = \Phi_L + \fnl\Phi_L^2 \, ,
\end{equation}
where the subscript $L$ denotes the dominant linear, Gaussian component. Since this expansion is local in the configuration space, this is called the ``local'' type non-Gaussianity. This parametrization is also widely used in the studies on the inflationary non-Gaussianity of the comoving curvature perturbation $\calR$ as~\cite{Maldacena:2002vr}
\begin{equation}\label{Rfnl}
\calR = \calR_L + \frac{3}{5}\fnl\calR_L^2 \, .
\end{equation}
This is based on the linear relation between $\Phi$ and $\calR$ in the large scale limit during matter domination
\begin{equation}\label{Lin-relation}
\Phi = \frac{3}{5}\calR \, .
\end{equation}

This requires, however, a great caution. Primordial non-Gaussianity is one realization of non-linearity of the primordial perturbation. Hence, (\ref{Rfnl}) being based on the linear relation (\ref{Lin-relation}) is {\em not} guaranteed to be valid. To estimate $\fnl$ and in turn the degree of non-Gaussianity properly, we have to reconsider (\ref{Rfnl}) to take into account the full second order evolution effects during matter domination. This is important to constrain the generation mechanism of the primordial perturbation. For example, it is believed that the detection of local type non-Gaussianity would rule out any single field inflation models~\cite{Creminelli:2004yq,single-field}. But this strong conclusion is based on the doubtful parametrization (\ref{Rfnl}). This possibility has been anticipated and studied in a few literatures~\cite{Bartolo:2004if,Maldacena:2002vr,Creminelli:2004yq,BMR,Gao:2010ti,MDeffect,Boubekeur:2009uk}, but they are incomplete. As we will show, in the literature, the non-linear evolution effects in $\fnl$ were derived often based on the linear Sachs-Wolfe relation. In order to estimate $\fnl$ properly, the CMB temperature fluctuations should include secondary effects as well~\cite{MDeffect,Boubekeur:2009uk}. Also, in the literature, the non-linear effects dominant on smaller scales have been neglected. With the upcoming observations, LSS provides another powerful probe to study non-linearities in cosmological perturbations. Thus, to constrain non-Gaussianity using LSS the non-linear effects on small scales are important.

In this article, we present the proper second order relation between $\Phi$ and $\calR$ generated by non-linear evolution during matter domination valid on all scales. With the correct prescription, $\fnl$ in $\Phi$ induces a number of additional second order terms to (\ref{Rfnl}). Further, starting from the primordial curvature perturbation generated during inflation, we show that the evolution effects give rise to sizable contributions to $\Phi$. Also we make a clear distinction between the Bardeen curvature $\Phi$ and the Bardeen potential $\Psi$, which is important beyond linear perturbation theory.

Our starting point is the metric with scalar-type perturbations~\cite{Bardeen-1988}
\begin{equation}
ds^2 = -(1+2\alpha)dt^2 - 2a\beta_{,i}dtdx^i + a^2 \left[ \left( 1 + 2\varphi \right)\delta_{ij} + 2\gamma_{,ij} \right] dx^idx^j \, ,
\end{equation}
where we have assumed flat background geometry and set $c=1$. We take $\gamma = 0$ as the spatial gauge (threading) condition. The comoving curvature perturbation $\calR$ and the Bardeen curvature $\Phi$, both gauge invariant, correspond to $\varphi-\varphi^2$ in the comoving gauge condition $v = 0$ with $v$ being the spatial component of the fluid four-vector $u_i = -av_{,i}$, and $\varphi$ in the zero-shear gauge condition $\beta = 0$, respectively. The reason why we define $\calR$ in such a way is because we want to separate the primordial non-linear component which is usually computed with the metric $g_{ij} = a^2e^{2\calR}\delta_{ij}$. We further introduce Bardeen potential $\Psi$, which is $\alpha$ in the zero-shear gauge. To linear order we have $\Psi = - \Phi$ but as we go to non-linear order it is important to distinguish between the Bardeen {\em curvature} $\Phi$ and the Bardeen {\em potential} $\Psi$. Note that to linear order and in the sub-horizon limit to second order $\Psi$ coincides with the Newtonian gravitational potential. We consider a pressureless fluid. Up to second order, assuming $\Lambda = 0$, the relatively growing mode exact solutions are~\cite{Boubekeur:2009uk,Noh:2003yg,Hwang:2007ni}
\begin{align}
\label{MDR}
\calR = & C - C^2- \frac{1}{5(aH)^2} \left[ \frac{1}{2}C^{,i}C_{,i} + \Delta^{-1}\left(C^{,i}\Delta{C}\right)_{,i} \right] \, ,
\\
\label{MDPhi}
\Phi = & \frac{3}{5}\calR + \frac{3}{25}\calR^2 + \frac{6}{25}\Delta^{-1} \left[ -\calR\Delta\calR + 3\Delta^{-1}\left(\calR\calR^{,ij} \right)_{,ij} \right] + \frac{9}{175(aH)^2} \left[ \frac{1}{2}\calR^{,i}\calR_{,i} - \Delta^{-1}\left(\calR^{,i}\Delta\calR\right)_{,i} \right] \, ,
\\
\label{MDPsi}
\Psi = & - \Phi + \frac{3}{25}\calR^2 + \frac{3}{5}\Delta^{-1} \left[ -\calR\Delta\calR + 3\Delta^{-1}\left(\calR\calR^{,ij} \right)_{,ij} \right] \, ,
\end{align}
where $C=C(\mathbi{x})$ is a constant coefficient of the relatively growing solution of $\varphi$ in the comoving gauge in the large scale limit. Compared with (\ref{Lin-relation}), (\ref{MDPhi}) is the correct relation between $\Phi$ and $\calR$ valid to second order perturbation. It shows clearly that in the matter dominated era we have non-trivial contributions at second order.

In the large scale limit $\calR$ is known to be conserved throughout the evolution even to second order~\cite{Hwang:2007ni,Lyth:2004gb}: $\calR$ is constant to second order for general time-varying equation of state or field potential. In the conventional scenario, $\calR$ on the super-horizon scales is generated from the quantum fluctuations during inflation but it may be coming from other mechanism. For our current purpose, it is sufficient to note that the large scale constant of (\ref{MDR}) represents the primordial component of the comoving curvature perturbation. From below we will write $C-C^2 \equiv \calR_\text{prim}$.

On the other hand, observations are often characterized by $\Phi$ or $\Psi$ which, to linear order or to second order in the small scale limit, satisfy $\Psi = - \Phi$ with $\Psi$ being the same as the perturbed Newtonian gravitational potential~\cite{Bardeen:1980kt}. In the context of the CMB temperature fluctuations, the observation is sensitive to the non-Gaussianity of $\Psi$. In the matter dominated era and in the large scale limit, with $\Lambda=0$ and flat geometry, we have~\cite{Bartolo:2004if}
\begin{equation}
\left. {\delta T \over T} \right|_O = \left. \left[ {1 \over 3} \Psi - {5 \over 18} \Psi^2 \right] \right|_E \, ,
\label{SW}
\end{equation}
where the subscripts $O$ and $E$ indicate the observed and emitted epochs, respectively.

To translate $\fnl$ given in terms of $\Phi$ in (\ref{PhifNL}) into $\calR$, we use (\ref{PhifNL}) and (\ref{MDPhi}) to find, in the large scale limit,
\begin{align}
\calR = & \calR_L + \frac{3}{5}\fnl\calR_L^2 - \frac{1}{5}\calR_L^2 + \frac{2}{5}\Delta^{-1} \left[ \calR_L\Delta\calR_L - 3\Delta^{-1}\left( \calR_L\calR_L^{,ij} \right)_{,ij} \right] \, .
\label{properRfNL}
\end{align}
Thus, the conventional parametrization in (\ref{Rfnl}) based on the linear relation misses substantial contributions. These terms describe the evolution effects after inflation up to matter dominated era.

With the proper non-linear relation between $\calR$ and $\Phi$, we now proceed to evaluate $\fnl$ correctly. In the Fourier space, we introduce non-local $\fnl^{(\Phi)}$ and $\fnl^{(\calR)}$ as
\begin{align}
\Phi(\mathbi{k}) = & \Phi_L(\mathbi{k}) + \int \frac{d^3q_1d^3q_2}{(2\pi)^3} \fnl^{(\Phi)}(\mathbi{q}_1,\mathbi{q}_2) \Phi_L(\mathbi{q}_1)\Phi_L(\mathbi{q}_2)\delta^{(3)}(\mathbi{k}-\mathbi{q}_{12}) \, ,
\label{nlPhi}
\\
\calR_\text{prim} (\mathbi{k}) = & \calR_L (\mathbi{k}) + \int \frac{d^3q_1d^3q_2}{(2\pi)^3} \frac{3}{5}\fnl^{(\calR)}(\mathbi{q}_1,\mathbi{q}_2) \calR_L(\mathbi{q}_1)\calR_L(\mathbi{q}_2) \delta^{(3)}(\mathbi{k}-\mathbi{q}_{12}) \, ,
\label{nlR}
\end{align}
where $\mathbi{q}_{12} \equiv \mathbi{q}_1+\mathbi{q}_2$. Here, $\fnl^{(\calR)}$ denotes the primordial non-Gaussianity, generated from for example inflation in the conventional scenario. From (\ref{MDR}) and (\ref{MDPhi}), we can derive
\begin{align}
\label{fNL-Phi}
\fnl^{(\Phi)}(\mathbi{k}_1,\mathbi{k}_2) = & \fnl^{(\calR)}(\mathbi{k}_1,\mathbi{k}_2) + \frac{2}{3}g(\mathbi{k}_1,\mathbi{k}_2) + \left(\frac{k_{12}}{aH}\right)^2 h(\mathbi{k}_1,\mathbi{k}_2) \, ,
\\
g(\mathbi{k}_1,\mathbi{k}_2) \equiv & \frac{1}{2} - \frac{k_1^2+k_2^2}{2k_{12}^2} + \frac{3}{2}\frac{\left(\mathbi{k}_1\cdot\mathbi{k}_{12}\right)^2 + \left(\mathbi{k}_2\cdot\mathbi{k}_{12}\right)^2}{k_{12}^4} \, ,
\\
h(\mathbi{k}_1,\mathbi{k}_2) \equiv & \frac{1}{21k_{12}^2} \left( 2 \mathbi{k}_1\cdot\mathbi{k}_2 + 5 \frac{k_1^2}{k_{12}^2} \mathbi{k}_2\cdot\mathbi{k}_{12} + 5 \frac{k_2^2}{k_{12}^2} \mathbi{k}_1\cdot\mathbi{k}_{12} \right) \, .
\end{align}
Apparently, the two non-local parameters $\fnl^{(\calR)}$ and $\fnl^{(\Phi)}$ are {\em not} the same. They are related in a non-trivial manner, with $\fnl^{(\Phi)}$ receiving non-linear evolution effects during matter domination. These effects are composed of two parts: one being  independent of the horizon scale $k_H=aH$, and the other being dominant on sub-horizon scales. Both are shape dependent. If we introduce $\fnl^{(\Psi)}$ similarly defined as in (\ref{nlPhi}) with $\Phi$ replaced by $\Psi$, by using the relations between $\calR$ and $\Psi$ presented in (\ref{MDR}) and (\ref{MDPsi}), we can find
\begin{equation}
\label{fNL-Psi}
\fnl^{(\Psi)}(\mathbi{k}_1,\mathbi{k}_2) = - \fnl^{(\calR)}(\mathbi{k}_1,\mathbi{k}_2) + g(\mathbi{k}_1,\mathbi{k}_2) - \frac{1}{2} - \left(\frac{k_{12}}{aH}\right)^2 h(\mathbi{k}_1,\mathbi{k}_2) \, .
\end{equation}
In the large scale limit this relation is presented in Refs.~\cite{Bartolo:2004if,BMR}. We will address this issue later. Both $\fnl^{(\Phi)}$ and $\fnl^{(\Psi)}$ exhibit similar structure as we can see from (\ref{fNL-Phi}) and (\ref{fNL-Psi}), and in the following we will closely analyze $\fnl^{(\Phi)}$.

We can compute $\fnl^{(\calR)}$ by adopting for example the cubic order action~\cite{Maldacena:2002vr,single-field,Gong:2011uw} or the $\delta{N}$ formalism~\cite{Lyth:2005fi} to calculate the primordial bispectrum. It is known in the literature that for {\em all} inflation models where only one degree of freedom is important during seed generation, in the squeezed limit we have the consistency relation $\fnl^{(\calR)} = 5(1 - n_s)/12$ with $n_s$ being the spectral index of the scalar power spectrum~\cite{Maldacena:2002vr,Creminelli:2004yq}. Thus, it is often claimed that the detection of $\left|\fnl\right| \gtrsim \calO(1)$ in the squeezed configuration will rule out all classes of single field inflationary models. But, as stressed before, this is based on ignoring the classical contribution we have studied: in (\ref{fNL-Phi}) we have shown pure classical contributions in matter dominated era which have far larger magnitude, so that tiny primordial $\fnl^{(\calR)}$ is completely subdominant compared with $\fnl^{(\Phi)}$.

\begin{figure}[t]
 \begin{center}
 \includegraphics[width=16cm]{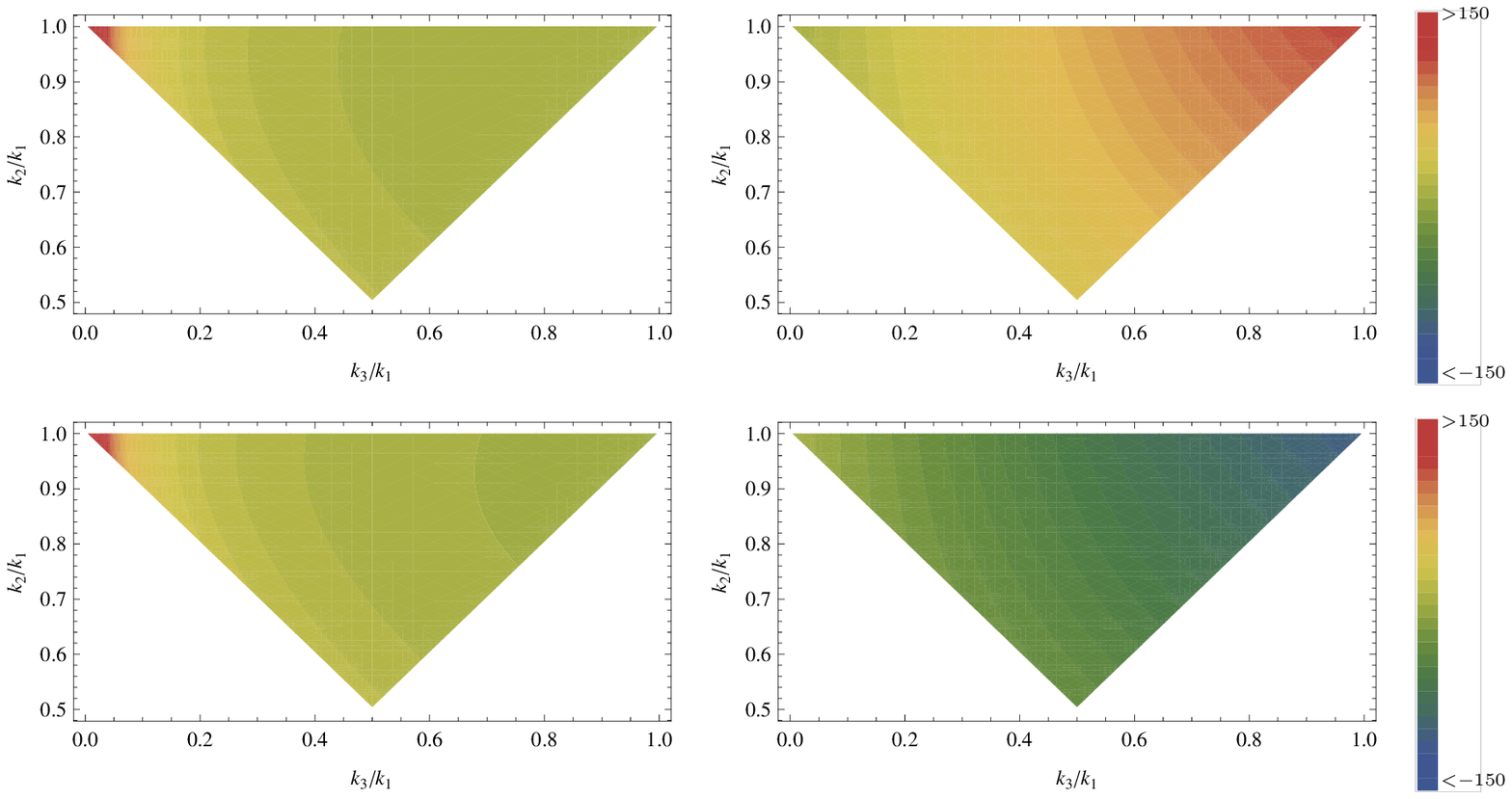}
 \end{center}
 \caption{Contour plot of the contribution to the shape function $(k_1k_2k_3)^2B_\Phi(\mathbi{k}_1,\mathbi{k}_2,\mathbi{k}_3)$
 of the horizon independent terms (upper-left), and the horizon dependent terms (upper-right). In the lower panel we show the corresponding shape function of $\Psi$. We can see that in the left panel the shape functions are peaked at the squeezed limit. This means some of the pure non-linear evolution effects during matter dominated era take place over all scales, even outside the horizon. Meanwhile, in the right panel, the shape functions show their maximum amplitudes at the equilateral limit. Since these terms are heavily suppressed on super-horizon scales, the causal non-linear effects on the sub-horizon scales are maximized when the three momenta are of more or less the same size. As we can see from (\ref{fNL-Phi}) and (\ref{fNL-Psi}), on large scales the shape functions of $\Psi$ and $\Phi$ are similar, and on small scales the magnitude is precisely the same but the sign is opposite. This reflects the second order relation $\Phi = -\Psi$ on small scales. We have set $k_1 = 10 k_H$ for the right panel.}
 \label{fig:shape}
\end{figure}

We can calculate the bispectrum of $\Phi$ as
\begin{equation}\label{bispectrum}
B_\Phi(\mathbi{k}_1,\mathbi{k}_2,\mathbi{k}_3) = 2\fnl^{(\Phi)}(-\mathbi{k}_1,-\mathbi{k}_2)P_\Phi(k_1)P_\Phi(k_2) + \text{(2 perm)} \, .
\end{equation}
Given $\Phi$ in (\ref{nlPhi}) and $\fnl^{(\Phi)}$ in (\ref{fNL-Phi}), we can estimate exactly the shape and the magnitude of the bispectrum. In Figure~\ref{fig:shape}, we present the dimensionless shape function $(k_1k_2k_3)^2B_\Phi(\mathbi{k}_1,\mathbi{k}_2,\mathbi{k}_3)$, normalized by the amplitude of the power spectrum $A_\Phi=P_\Phi(k)/k^{n_s-4}$, and that of $\Psi$, apart from possibly negligible $\fnl^{(\calR)}$. We present two contributions separately: (left) the horizon independent terms, and (right) the horizon dependent terms. We can see that the former is peaked in the squeezed limit, while the latter in the equilateral limit. Since $\fnl^{(\Phi)}$ is explicitly momentum dependent, in fact the perfect local non-Gaussinity ansatz, i.e. $\fnl$ is a constant, does not work. This suggests that $\fnl$ is, despite of its popularity, not a good parameter to describe the bispectrum. Nevertheless, we can estimate $\fnl$ by comparing (\ref{bispectrum}) with the bispectrum we can find from the perfect local ansatz (\ref{PhifNL}). Then we find the local $\fnl$ in the large scale limit as, with $\fnl^{(\calR)}$ being ignored, $\fnl = \fnl\left(\mathbi{k}_1,\mathbi{k}_2,\mathbi{k}_3\right) $ which depends on the shape of the triangle. It gives $\fnl = 1$ and $1/6$ in the squeezed and equilateral limits, respectively. But in the folded limit $\fnl$ is also dependent on $n_s$, and gives $19/15$ for $n_s=1$. Therefore, we have
\begin{equation}
\frac{1}{6} \leq \fnl \lesssim \frac{19}{15} \, .
\end{equation}
In terms of $\fnl^{(\Psi)}$, in the same way we have $\fnl = 1$, $-1/4$ and $7/5$ in the squeezed, equilateral and folded limits, respectively. Therefore, in terms of $\Psi$ we have
\begin{equation}
- \frac{1}{4} \leq \fnl \lesssim \frac{7}{5} \, .
\end{equation}
Thus, the detection of local $\fnl$ with the amplitude of $\calO(1)$ does not necessarily exclude single field inflation models. From another perspective, (\ref{fNL-Phi}) and (\ref{fNL-Psi}) predict that, unless primordial $\fnl^{(\calR)}$ is bigger than $\calO(1)$, we should find the classical contributions to $\fnl$ as a consequence of general relativity in matter dominated era.

\begin{figure}[t]
 \begin{center}
  \includegraphics[width=15cm]{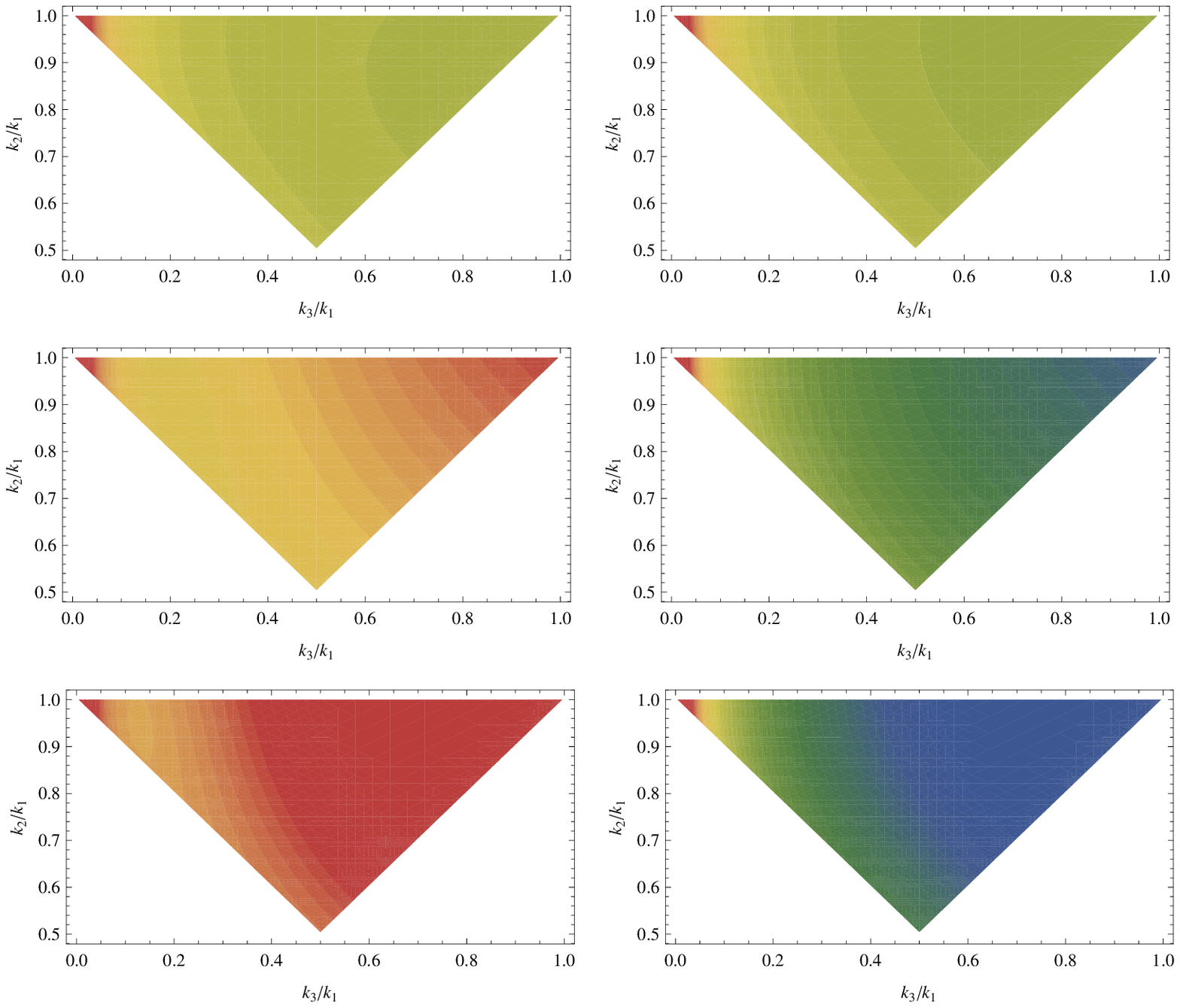}
 \end{center}
 \caption{Contour plot of the full shape functions for  (left) $\Phi$ and  (right) $\Psi$ with (top) $k_1 = 2k_H$, (middle) $k_1 = 10k_H$ and (bottom) $k_1 = 15k_H$. Depending on the scale we probe, we may have a mixed shape of the squeezed and equilateral limits. On smaller scales, the causal, sub-horizon gravitational interactions become important and as a result we find a larger peak at the equilateral limit. The contour scaling is the same as Figure~\ref{fig:shape}.}
 \label{fig:scale}
\end{figure}

On smaller scales relevant for LSS observations the horizon dependent terms dominate and we have non-trivial behaviour of $\fnl^{(\Phi)}$ and $\fnl^{(\Psi)}$. In Figure~\ref{fig:scale}, we present the total shape functions for  (left) $\Phi$ and (right) $\Psi$ with different size of $k_1$: (top) $k_1 = 2k_H$, (middle) $k_1 = 10k_H$ and (bottom) $k_1 = 15k_H$, respectively. As one may expect, as we probe smaller scales, the horizon dependent contributions become more dominant, giving rise to a larger equilateral peak. This suggests that we have another probe of the general relativistic effects on smaller scales, where for example the scale dependent bias may serve as a powerful probe of the primordial non-Gaussianity~\cite{nGbias}.

Here, we compare our result with the previous studies. In Refs.~\cite{BMR,Gao:2010ti}, on large scales the non-linear evolution effects contribute to $\fnl$ as
\begin{equation}\label{fnl_BMR}
\fnl = \fnl^{(\calR)} - g + \frac{4}{3} \, .
\end{equation}
This obviously leads to different shape of the bispectrum as well as magnitude. For example, in the squeezed limit from (\ref{fnl_BMR}) we find $\fnl = -1/6$. This discrepancy arises because (\ref{fnl_BMR}) is based on the {\it linear} Sachs-Wolfe relation. Indeed, if we assume that the linear component of (\ref{SW}) holds non-linearly as $\delta{T}/T = -\Phi/3$ and apply the exact solutions (\ref{MDPhi}) and (\ref{MDPsi}) to find $\fnl$, we can obtain (\ref{fnl_BMR}).

In order to correctly incorporate the full second order evolution effects we must also take into account other secondary effects, such as the integrated Sachs-Wolfe (ISW) effects, weak lensing, and so on. If $\fnl$ is confirmed to be $\calO(1)$ so that the primordial contribution $\fnl^{(\calR)}$ is very small, it is very important to correctly identify the non-linear, second order effects. We would like to stress that apart from the second order transfer function, ISW, lensing etc, the correct relation between $\calR$ and $\Phi$ or between $\calR$ and $\Psi$ should be the starting point.

There is another issue we should comment. As we mentioned before, at linear order and in the sub-horizon limit at second order, we have $\Phi = -\Psi$, and $\Psi$ coincides with perturbed Newtonian gravitational potential. Thus, in these two cases, up to sign convention, practically it does not matter which variable we choose to define the non-linear parameter $\fnl$. In general, however, at proper second order this is no longer the case, as we have shown in (\ref{MDPhi}) and (\ref{MDPsi}). On small scales, where LSS serves as a powerful probe of non-Gaussianity, we use the Poisson-like relation between the matter density perturbation $\delta$ and the Bardeen curvature $\Phi$,
\begin{equation}\label{Poisson}
\delta(k) = \mathcal{M}(k)\Phi(k) \, ,
\end{equation}
where $\mathcal{M}(k)$ is a combination of matter transfer function, window function and so on. To study the effects of non-Gaussianity, we substitute the local ansatz (\ref{PhifNL}) into $\Phi$ in (\ref{Poisson}). In Newtonian context, $\Phi$ should be in fact $-\Psi$, the gravitational potential, but even in the context of general relativity (\ref{Poisson}) is also valid in the two cases mentioned above. However, at second order where we can properly consider non-linearity and thus estimate non-Gaussianity, especially in the large scale, this is no longer the case as we have shown in this work.

In this article, we have reconsidered the relation between $\calR$, $\Phi$ and $\Psi$. The widely used relation (\ref{Rfnl}) is properly extended to second order as (\ref{MDPhi}) and (\ref{MDPsi}). Using the correct second order relation between $\calR$, $\Phi$ and $\Psi$, we have explicitly clarified the relation between the non-linear parameters in (\ref{fNL-Phi}) and (\ref{fNL-Psi}): $\calR$ is generated, in the conventional scenario, during inflation and the the Bardeen potential $\Psi$ is directly related to the temperature fluctuations as (\ref{SW}) and the Newtonian gravitational potential. The observationally relevant $\fnl$ contains substantial non-linear evolution effects during matter dominated era besides the primordial contribution. While the one effective on super-horizon scales gives the maximum amplitude of the bispectrum at the squeezed limit, the one dominant on sub-horizon scales gives an equilateral peak. This suggests, in addition to our main findings in this article, another interesting way of probing general relativistic effects using non-Gaussianity on different scales.

\subsection*{Acknowledgements}

We thank Nicola Bartolo, Eiichiro Komatsu, Antonio Riotto and Takahiro Tanaka for useful conversations and correspondences.
We acknowledge the workshop ``WKYC 2011 -- Future of Large Scale Structure Formation'' at Korea Institute for Advanced Study where this work was initiated.
JG was supported in part by a Korean-CERN fellowship.
JH was supported by Korea Research Foundation Grant funded by the Korean Government (KRF-2008-341-C00022).
HN was supported by Mid-career Research Program through National Research Foundation funded by the MEST (No. 2011-0000054).

\end{document}